\documentclass[12pt,a4paper,final]{iopart}

\usepackage{iopams}  
\usepackage{graphicx}
\usepackage[breaklinks=true,colorlinks=true,linkcolor=blue,urlcolor=blue,citecolor=blue]{hyperref}
\usepackage[TS1,T1]{fontenc}
\usepackage{textcomp}
\begin{document}

\title[Estimating dynamic mechanical quantities]{Estimating dynamic mechanical quantities and their associated uncertainties: application guidance}

\author{Trevor Esward}
\address{National Physical Laboratory, Teddington TW11 0LW, UK}
\eads{\mailto{trevor.esward:@npl.co.uk}}

\author[cor1]{Sascha Eichst\"adt}
\address{Physikalisch-Technische Bundesanstalt, Berlin, Germany}
\ead{sascha.eichstaet@ptb.de}

\author{Ian Smith}
\address{National Physical Laboratory, Teddington TW11 0LW, UK}
\ead{ian.smith@npl.co.uk}

\author{Thomas Bruns}
\address{Physikalisch-Technische Bundesanstalt, Berlin, Germany}
\ead{thomas.bruns@ptb.de}

\author{Peter Davis}
\address{National Physical Laboratory, Teddington TW11 0LW, UK}
\ead{peter.davis@npl.co.uk}

\author{Peter Harris}
\address{National Physical Laboratory, Teddington TW11 0LW, UK}
\ead{peter.harris@npl.co.uk}

\begin{abstract}

Recently several European National Measurement Institutes have established traceable calibration methods for dynamic mechanical quantities, e.g., dynamic force, torque and pressure. However, the use in industry and elsewhere of dynamic calibration information provided on certificates is not straightforward. Typically it is necessary to employ deconvolution techniques to obtain estimates of measurands, and the deconvolution method itself and the associated algorithms are sources of uncertainty that must be included in uncertainty budgets. There is a need for practical guidance for end users on how to use the newly-available dynamic calibration information. To this end we set out an approach to the evaluation of uncertainties associated with dynamic measurements that we believe covers the most relevant cases. The methods have been embodied in publicly-available software and we show how they can be used to tackle some example problems. We believe that the methods lead to more reliable estimates of the relevant measurands and their associated uncertainties.

\end{abstract}


\section{Introduction}

Many applications of the measurement of quantities such as force, torque and pressure are dynamic, i.e., the measurand shows a strong variation over time. Transducers are in most cases calibrated by static procedures owing to a lack of commonly accepted procedures and documentary standards that take account of their dynamic behaviour. It is well known that mechanical sensors exhibit a behaviour that shows an increasing deviation from static sensitivity characteristics as the frequency content of the measurand increases. The lack of standards for dynamic calibration also applies to the electrical conditioning components of the measurement chain.

European collaborative projects~\cite{IND09,14SIP08} provided outputs in the forms of general dynamic models for the complete calibration measurement chain, methods for uncertainty evaluation in line with uncertainty evaluation for static measurements, and general procedures for correcting measurements for dynamic effects. However, these outputs have not yet been embodied in documentary standards and international guidance documents or in software that can be used in industrial applications to correct measurements and provide uncertainty evaluations that are compliant with the 'Guide to the expression of uncertainty in measurement' (GUM)~\cite{JCGM100}.

Calibration certificates and associated information provided for dynamic quantities by National Measurement Institutes (NMIs) and accredited calibration laboratories can take several forms, such as parameterized models of the sensors and measuring systems that are calibrated, or frequency response data that describe the amplitude and phase of the output of a calibrated system as a function of frequency in comparison to the input to the system. In addition, sensors alone may be calibrated, so that the end user has to understand how the remainder of the measuring system (amplifiers, filters, digital acquisition systems, etc.) affects the performance of the calibrated system.

The calibration methods may also be based on a variety of input signals, such as sine waves, chirps, steps and impulses, and the choice of signal determines what calibration information may be obtainable and how it may be used. Therefore, industrial end users require guidance on what calibration information to request from NMIs and accredited calibration laboratories, guidance on how to use this information in their own dynamic measurement applications to ensure compliance with the GUM, and software that demonstrates the guidance in action. The aim of this paper is to provide practical guidance on the evaluation of measurement uncertainty in dynamic applications and to identify software, mathematical tools and documents that are of assistance in this task. Readers who require an introduction to deconvolution methods in metrology are referred to~\cite{EicElsEswHes10}. 

\section{Definitions relevant to dynamic measurements}

A quantity is called a dynamic quantity when its value depends on time. Similarly, quantities that depend on wavelength, spatial coordinates or temperature, for instance, can be treated as `dynamic'.

Typically, in a model of a dynamic measurement at least one of the input quantities and the measurand are dynamic quantities. An important aspect is that the relation between the dynamic quantities is given by a dynamic model based on differential equations rather than algebraic equations. Some models of dynamic measurements can, however, be cast as multivariate measurement models that can be handled with GUM Supplement 2~\cite{JCGM100,JCGM102}. Note that the time-dependence can be continuous or discrete. The measured values of a dynamic quantity for different time instants are generally not independent, resulting in a non-zero auto-correlation that has to be taken into account for the evaluation of uncertainty. The class of linear time-invariant (LTI) systems is the simplest class of dynamic system models, see \cite{OppSch89} for details. It is appropriate for a wide range of applications and all the examples discussed in this paper are modelled as LTI systems.

\section{Example of a typical dynamic measurement problem}

The acceleration of a specimen while exposed to a force that varies in time is a dynamic quantity, and the measurement of this acceleration is a dynamic measurement.

\begin{figure}[htp]
\centerline{\includegraphics[height=80mm,width=120mm]{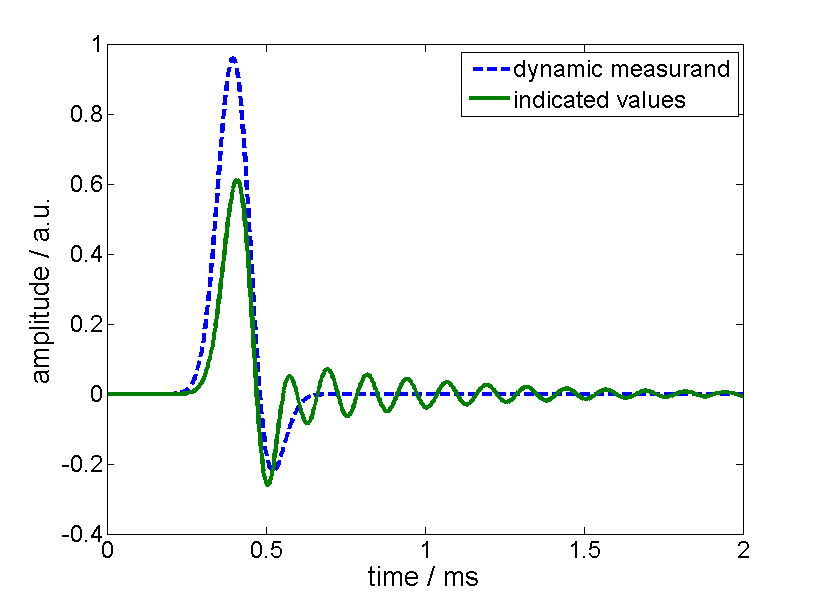}}
\caption{\small Simulated dynamic measurement illustrating the effect of a frequency-dependent behaviour of the measuring system.}
\label{fig: dynamic ringing}
\end{figure}

Suppose that an accelerometer, which has a resonance frequency, is used to measure an acceleration that has appreciable frequency content in the vicinity of the resonance frequency. The values indicated by the measuring system then show a ringing effect resulting in a time-varying deviation from the dynamic measurand. Figure \ref{fig: dynamic ringing} shows such an effect where the low-frequency behaviour of the simulated sensor contributes mainly to the reduced peak height whereas the resonance frequency at 8\,kHz leads to the characteristic ringing in the indicated values after the main signal. After a scaling and a time-shift of the indicated values, appreciable estimation errors would remain. As a result of the interaction between the measurand and the accelerometer that arises from its resonant behaviour a dynamic measurement model is needed.

The taking into account of the effects of insufficient measurement bandwidth requires a correction (or deconvolution) process to be applied to the measuring system outputs as part of the estimation of the dynamic measurand. This correction is exactly analogous to correcting for systematic effects in static or steady-state measurements, and in both the dynamic and static cases it is necessary to evaluate the uncertainty appropriate to the correction. Later in this paper we discuss in detail a high intensity shock calibration problem that demonstrates many of the challenges that arise in the metrology of dynamic systems~\cite{Bruns2017}. 

\section{Dynamic measurements for which the GUM can be used directly}

There are many cases in which a quantity depends on time and the conventional methodology of the GUM may be applied.  

Consider a sinusoidally varying signal of the form \begin{equation} \label{dynamic_sine} Y(t)= A\sin(\omega t + \phi),\end{equation}
where $\omega$ denotes angular frequency, $\phi$ denotes phase angle and $A$ is a constant denoting amplitude. Knowing that the signal takes a sinusoidal form and knowing the amplitude, phase and frequency means that the signal is known for any and all times that may be of interest. Furthermore, one may analyse such a time-varying signal in the frequency domain rather than in the time domain by the use of the Fourier transform, the output of which gives the values of these parameters directly, provided the sensitivity of the measuring system at the frequency of interest is known. Note, too, that this approach is valid for both continuous and discrete representations and realizations of the signal as equation~(\ref{dynamic_sine}) applies in both cases, i.e., $Y(t)$ may be evaluated at any required time and the Fourier transform then becomes a discrete Fourier transform. 

The effects of noise can also be taken into account. Suppose the signal of interest is in the form of a single frequency sinusoidally varying signal as in equation~(\ref{dynamic_sine}) and the measured signal is \[ X(t) = A\sin(\omega t + \phi) + \epsilon(t),\] where $\epsilon(t)$ is noise. The effect of the noise is such that the Fourier transform (or discrete Fourier transform in the discrete case) of the measured signal will contain other frequency components, and the estimates of amplitude, phase and frequency obtained from the Fourier transform will be perturbed from the values of these parameters for the signal of interest.

The above discussion can be extended to the case of a signal of arbitrary complexity, provided that the purpose of the measurement is to estimate the amplitude and phase of a single sinusoidal component of known frequency. Instruments in common use throughout metrology, such as the lock-in amplifier (or phase-sensitive detector), are designed specifically to extract the parameters of a signal component at a specific frequency from measured signals that contain many frequency components and much noise. Once again, conventional GUM methodologies for uncertainty evaluation can be applied to such cases. For further information concerning the use of lock-in amplifiers in uncertainty evaluation, see~\cite{Clarkson2010}.

The examples considered above are all instances of dynamic measurements but applied to single frequency signals or to extract single or narrowband frequency information from complex signals. They demonstrate that the observation of a time-varying signal does not necessarily require the development of a dynamic rather than a simple algebraic model of the measurement. The same holds true for any other dynamic measurement where a parametric model for the signal is available, and the estimation then becomes a regression task.

There is a further consideration that may mean that although the signal of interest is dynamic and has a broad frequency content, a dynamic model may not be needed. This is the case when the measuring system has a bandwidth that contains that of the signal of interest and the frequency response of the measuring system for the frequencies contained in the signal can be regarded as constant.  Then, for the analysis in this case a single value for the system's sensitivity is sufficient (with its own uncertainty that may have been determined by a calibration process), which acts as a multiplicative factor to be applied to the observed output signal, and GUM methodologies appropriate for static cases can be applied, provided that the signals are either discrete-time or depend on a finite number of parameters. In both cases, multivariate uncertainty evaluation methods, i.e. GUM Supplement 2, can be applied.  

A measurand, or a signal that contains the measurand of interest, may be time-varying. However, it is the interaction between the time-varying measurand and the measuring system that determines whether a dynamic measurement model is needed or not. Thus, a key aspect of dynamic measurement is the choice of measuring system. If the bandwidth of the measuring system is sufficient for the measurement task, uncertainty evaluation is simplified. Careful selection of the measuring system and its components by the metrologist may avoid the need to consider dynamic measurement models of the kind described later in this paper. 

Of course, it may be the case that the purpose of a measurement is not to estimate the complete time history of the signal of interest, and that the measurand is some feature of the signal such as a maximum or minimum value, the time at which a maximum or minimum occurs, or some average quantity of the signal such as its root mean square value. In such cases it is possible that to obtain the best estimate of such a measurand the reconstruction of the complete signal is necessary even if this is not the purpose of the measurement, so that the methods described later in this paper are needed. 

\section{Dynamic measurement models}

\subsection{Typical workflow for a dynamic measurement}

\begin{figure}[htp]
\centerline{\includegraphics[width=0.8\columnwidth]{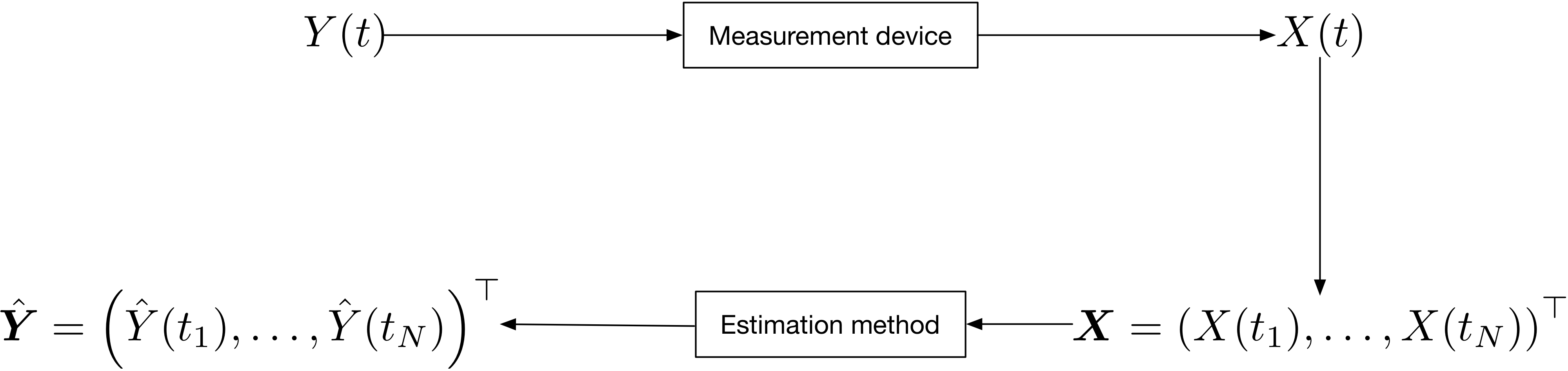}}
\caption{\small Illustration of the typical workflow in the analysis of a dynamic measurement. The top part is carried out in the continuous time domain whereas the bottom part is in the discrete time domain.}
\label{fig: dynamic workflow}
\end{figure}

We consider the direct measurement of a dynamic measurand with a measurement device. The corresponding typical workflow in a dynamic measurement is illustrated in Figure~\ref{fig: dynamic workflow}. The continuous time-varying $Y(t)$ is the input to the measurement device with continuous time-varying output $X(t)$. Note that the dynamic quantities $Y(t)$ and $X(t)$ can also be multivariate. It  is assumed that an analogue-to-digital conversion of the dynamic system output $X(t)$ results in an equidistant discrete-time dynamic quantity ${\bi X}=(X(t_1),\ldots,X(t_N))^\top$. The corresponding dynamic measurand is the discrete-time dynamic quantity ${\bi Y}=(Y(t_1),\ldots,Y(t_N))^\top$, an estimate of which is denoted by $\widehat{\bi Y} = (\widehat{Y}(t_1),\ldots,\widehat{Y}(t_N))^\top$.

\subsection{Mathematical model of the measurement device}
\label{sec:system model}

The mathematical model that describes the relation between the dynamic measurand $Y(t)$, which acts as input to the measurement device, and the corresponding dynamic indication quantity $X(t)$ is called a \emph{dynamic system model} or dynamic system and is denoted as
\begin{equation} 
X(t) = \mathcal{H}\left[Y(t)\right] . 
\label{Direct dynamic measurement model}
\end{equation}
The dynamic system model is part of the observation model and should not be confused with the measurement model, which describes the relation between an estimate $\widehat{\bi Y}$ of the discrete-time dynamic quantity $\bi Y$ and values of the discrete-time dynamic quantity $\bi X$ (as in Figure~\ref{fig: dynamic workflow}). However, the evaluation of the uncertainty associated with the estimate $\widehat{\bi Y}$ should include a contribution that recognizes that the behaviour of the measurement device will not be known exactly and the process of correcting the indications of the measurement device to account for the behaviour of the device will also be imperfect -- see equation~(\ref{deconvolution filter}) below.

The behaviour of an accelerometer, for example, can be modelled mathematically by an ordinary differential equation (ODE) with constant coefficients:
\begin{equation}
\ddot{X}(t) + 2\delta\omega_0\dot{X}(t) + \omega_0^2 X(t) = \rho A(t) \, ,
\label{ODE spring-damper model}
\end{equation}
where $\delta$ denotes a damping coefficient, $\omega_0=2\pi f_0$ the resonance frequency, $\rho$ a proportionality constant, and $\dot{X}(t)$ and $\ddot{X}(t)$ are, respectively, the first and second order derivatives of $X(t)$ with respect to $t$. The time-varying $A(t) [\equiv Y(t)]$ denotes the acceleration to which the sensor is exposed and $X(t)$ the corresponding time-varying values indicated by the measurement device.

Using the Laplace transform, the ODE~(\ref{ODE spring-damper model}) can be transformed to the system transfer function
\begin{equation}
H(s) = \frac{\rho}{s^2 + 2\delta\omega_0 s + \omega_0^2} \, ,
\label{LTI transfer function}
\end{equation}
with parameters $\delta,\omega_0$ and $\rho$, where $s$ is a complex-valued frequency parameter. The inverse Laplace transform of $H(s)$ is the system impulse response function $h(t)$. The three variants of the mathematical description of the accelerometer behaviour -- the ODE, the transfer function and the impulse response function -- are equivalent representations of the system model~(\ref{Direct dynamic measurement model}). (In the signal processing literature the system transfer function is typically denoted by a capital letter whereas the impulse response function is denoted by a lowercase letter.) The Laplace transform representation of the transfer function $H(s)$ is useful, as the substitution of ${\rm j}\omega$ for $s$ in equation~(\ref{LTI transfer function}) allows one to evaluate directly the amplitude and phase responses of the measuring system, which together define the frequency response.

A metrologist who carries out or requests a calibration of a measuring system is in practice determining the system model. The mathematical form of the system model defines which parameters (and their associated uncertainties) are required to be estimated as the outputs of the calibration process. For example, in the case of the accelerometer described by the ODE (\ref{ODE spring-damper model}), the calibration might return estimates of the parameters $\delta,\omega_0$ and $\rho$ together with information about the uncertainties associated with the estimates. Typically, however, the results of calibrations of sensors used in dynamic applications are presented as amplitude and phase information as a function of frequency, often at discrete frequencies, rather than as a continuous function of frequency. In such cases the calibration delivers only an approximation to the system model. A continuous function of frequency may be obtained by curve fitting to the discrete frequency data.

\subsection{Correlation in dynamic systems}
\label{A straightforward way to handle the dynamic measurement model is to select an increasing set of times}

Given an increasing set of times $t_1, \ldots, t_N$, consider measurement models of the form
\[
\eqalign{
  Y_1 & = f_1(X_1),\cr
  Y_2 & = f_2(X_1,X_2),\cr
      & \vdots \cr
  Y_N & = f_N(X_1,\ldots,X_N),\cr
}
\]
where $X_i$ and $Y_i$ denote, respectively, $X(t_i)$ and $Y(t_i)$, and $f_i$ expresses the dependence of $Y_i$ on $X_1,\ldots,X_i$ and caters for any latency between the stimulus to the measurement device and its response to that stimulus. The $i$th of these expressions can be used to provide the best estimate of $Y_i$ given the best estimates of $X_1,\ldots,X_i$.

The uncertainty considerations are a little more complicated. Given the covariance matrix associated with the best estimates of $X_i,\ i = 1,\ldots,N$, comprising the variance of each estimate and covariance between each pair of estimates, and using linearisations of the functions $f_i$, the covariance matrix associated with the best estimates of $Y_i$ can be evaluated according to GUM Supplement 2 \cite{JCGM102}. The measured values of a dynamic quantity for different time instants are generally not independent, which results in non-zero covariances and the covariance matrix for the best estimates of $X_i$ is not diagonal. Furthermore, the use of a broadband sensor to measure a broadband signal inevitably introduces correlation as all frequency components of the signal are processed by the same sensor so that individual frequency components are not independent. It is a matter of judgement for the metrologist as to whether such correlations need to be taken into account when estimating the measurand. 

\paragraph{Example: temperature in a room monitored by a thermometer} 

The temperature $X(t)$ in a room is monitored at discrete time instants $t_i$ by a thermometer mounted centrally. The correlation of the temperature measured at different instants due to the use of the same thermometer has to be taken into account. Consider the $t_i$ to be at a constant spacing. Then, values $X_i = X(t_i)$ and $X_{i+1} = X(t_{i+1})$, adjacent in time, would be expected to be highly correlated and indeed any pair from $X_1, \ldots, X_N$ would have some associated correlation. Denote by $u[X(t)]$ the standard uncertainty associated with $X(t)$ at instant $t$ and $u[X(t), X(t')]$ the covariance associated with $X(t)$ and $X(t')$ at instants $t$ and $t'$. This information would be obtained from knowledge of the specific measurement. Uncertainties and covariances associated with $X_1, \ldots, X_N$ would be obtained from
\[
\eqalign{
  u(X_i) &= u[X(t_i)], \quad i = 1, \ldots, N, \cr
  u(X_i, X_j) &= u[X(t_i), X(t_j)], \quad i = 1, \ldots, N, \quad j = 1, \ldots, N \quad (j \neq i).
}
\]
When $Y_i$ is represented by an additive measurement model in terms of $X_1, \ldots, X_N$, methods from GUM Supplement 2 \cite{JCGM102} can be used to propagate the uncertainties and covariances associated with $X_1, \ldots, X_N$ to obtain uncertainties and covariances associated with $Y_1, \ldots, Y_N$.

For example, suppose the measurand is the mean temperature over a certain time period and thus involves an integral of the time-varying temperature $X(t)$. The evaluation of uncertainty for a quadrature formula used as an approximation to the integral can be carried out as follows. The mean temperature $Y(t)$ from time $a$ to time $t$ is calculated from the time-varying function $X(t)$ by the integral
\[ Y(t) = \frac{1}{t-a}\int_{a}^t X(\tau) \, \rm d\tau .\]
Represent $X(t)$ by its values $X_1, \ldots, X_N$ at times $t_1, \ldots, t_N$, where $t_1 = a$, with a constant time spacing $\Delta t =  t_{i+1} - t_i$. Correspondingly, represent $Y(t)$ by its values $Y_1, \ldots, Y_N$ and assume that $\Delta t$ is sufficiently small that the integral can be adequately approximated using the trapezoidal rule. The resulting measurement model is then given by
\begin{equation}
\label{Multivariate measurement model for dynamic measurement}
\eqalign{
  Y_1 &= 0, \cr
  Y_2 &= \frac{1}{2}\Delta t\left(X_1 + X_2\right),  \cr
  Y_j &= \Delta t \left(\frac{1}{2}X_1 + X_2 + \cdots + X_{j-1} + \frac{1}{2}X_j \right), \qquad j = 3, \ldots, N.  
}
\end{equation}
Note that with this approach instead of calculating the arithmetic mean of the measured values $X_i$, the trapezoidal rule is employed to evaluate the above integral.  

Given best estimates of $X_1, \ldots, X_N$, regarded as the input quantities, evaluation of expressions (\ref{Multivariate measurement model for dynamic measurement}) yield best estimates of the output quantities $Y_1, \ldots, Y_N$. By providing standard uncertainties and covariances associated with $X_1, \ldots, X_N$, GUM Supplement 2 \cite{JCGM102} can be applied to evaluate the standard uncertainties and covariances associated with $Y_1, \ldots, Y_N$.

\subsection{Treatment of continuous models}

Often, a continuous dynamic system model takes the form of a differential equation:
\begin{equation}
\dot Y(t) = F[t, X(t), Y(t), \mathit \Gamma_1, \ldots, \mathit \Gamma_M],
\label{differential dynamic measurement model}
\end{equation} 
where the $\Gamma_i$ denote quantities, such as model parameters, that do not depend on time. Note that equation (\ref{differential dynamic measurement model}) can represent a single first-order differential equation as well as a system of first-order differential equations.

Stochastic processes are associated with the state of knowledge about the dynamic quantities $X(t)$ and $Y(t)$. Given the stochastic process $X_t$ that models the state of knowledge about the dynamic input quantity $X(t)$, the sought stochastic process $Y_t$ for the measurand $Y(t)$ is determined by solving the stochastic differential equation~(\ref{differential dynamic measurement model}). Obtaining the solution requires stochastic modelling and is beyond the scope of the GUM. When the dynamic quantities can be represented by discrete-time sequences, and when the model (\ref{differential dynamic measurement model}) can be replaced by a corresponding discrete model, propagation of uncertainties can be carried out in accordance with the GUM \cite{Eic12}.

For example, assume that the relation between a time-varying force $F(t) = m A(t)$ and one-dimensional, time-varying elastic deformation $Y(t)$ of a specimen can be modelled by the ordinary differential equation
\begin{equation}
\ddot{Y}(t) + 2\delta\omega_0 \dot{Y}(t) + \omega_0^2 Y(t) = m A(t) \, .
\label{forward ODE example}
\end{equation}
Note that equation (\ref{forward ODE example}) can be easily transformed into a system of first-order ODEs. In contrast to the example in \ref{sec:system model} we now consider the ODE as our measurement model. That is, we assume that the acceleration $A(t) [\equiv X(t)]$ is measured at equidistant discrete time instants $t_i$ for $i=1,\ldots,N$, and the measurand is the elastic deformation $Y(t)$ at these time instants. The ODE parameters $\Gamma_1,\Gamma_2,\Gamma_3$, being $\delta$, $\omega_0$ and the mass $m$, are assumed to be known. Under certain conditions the discrete-time measurements of $A(t)$ can be used to define a stochastic process $A_t$ as a model for the state of knowledge about the values of the continuous-time $A(t)$~\cite{Eic12}. Propagation of uncertainty would then require to solve the induced stochastic differential equation using stochastic calculus. The evaluation of uncertainty can be carried out in accordance with GUM Supplement~2 provided that the ordinary differential equation can be discretized using, for instance, finite differences.

\section{Linear time-invariant systems}

\subsection{System models for linear time-invariant systems}

A system is called \emph{linear} when the dynamic system model is linear in its dynamic inputs, i.e., for dynamic quantities $Y_1(t)$ and $Y_2(t)$ with real-valued scaling factors $c_1$ and $c_2$ it holds that
\[ \mathcal{H}\left[c_1Y_1(t)+c_2Y_2(t)\right] = c_1\mathcal{H}\left[Y_1(t)\right] + c_2\mathcal{H}\left[Y_2(t)\right] .\]
The linearity of the system in the dynamic quantities should not be confused with linearity in the system parameters. The sensor whose input-output relation is modelled by the transfer function (\ref{LTI transfer function}) is linear with respect to the input acceleration, but the system model depends non-linearly on the parameters $\delta,\omega_0$ and $\rho$. 

A system is called \emph{time-invariant} when the dynamic system model does not change with time, i.e., provided that if
\[ \mathcal{H}\left[Y(t)\right] = X(t), \]
a time shift in $Y(t)$ results in the same time shift in $X(t)$:
\[ \mathcal{H}\left[Y(t-t_0)\right] = X(t-t_0) .\]
The ODE system model~(\ref{ODE spring-damper model}) is a linear time-invariant (LTI) system. The class of LTI systems is the simplest class of dynamic system models. It is appropriate for a wide range of applications.

For an LTI system with impulse response function $h(t)$, the relation between the system input $Y(t)$ and system output $X(t)$ is given by the convolution equation~\cite{OppSch89}
\begin{equation}
X(t) = (h \ast Y)(t) = \int_{-\infty}^\infty h(t-\tau)Y(\tau) \, \rm d\tau .
\label{LTI: convolution}
\end{equation}
Various other equivalent forms exist to model the input-output-relation for LTI systems. For instance, the relation between system input and system output can be modeled by a linear state-space system model with system matrices ${\bi C}, {\bi D}, {\bi E}$ and $\bi F$:
\begin{equation}
\begin{array}{lcl}
	\dot{\bi Z}(t) &=& {\bi C}{\bi Z}(t) + {\bi D}{\bi Y}(t), \\
	    {\bi X}(t) &=& {\bi E}{\bi Z}(t) + {\bi F}{\bi Y}(t).
\end{array}
\label{LTI: state-space}
\end{equation}
The state-space model~(\ref{LTI: state-space}) is typically used for systems with multiple system inputs and outputs or for networks of dynamic systems. For single-input-single-output systems the state-space model (\ref{LTI: state-space}) can be transformed to a transfer function model under certain regularity conditions~\cite{OppSch89}.

Neither the convolution equation~(\ref{LTI: convolution}) nor the state-space system model~(\ref{LTI: state-space}) are measurement models. Instead, the estimation of the value of the measurand requires the solution of an inverse problem. 

\subsection{Measurement models for linear time-invariant systems}

Equation (\ref{LTI: convolution}) shows how the system output $X(t)$ of the LTI system depends on the measurand $Y(t)$ through $h(t)$. In order to recover $Y(t)$ from $X(t)$, ideally, we construct a second LTI system with impulse response function $g(t)$ such that
\[
(g \ast X)(t) = [g \ast (h \ast Y)](t) = Y(t),
\]
so that applying the second LTI system to $X(t)$ recovers $Y(t)$. In practice, the second LTI system is implemented as a digital deconvolution filter, i.e., a digital filter designed such that the input to the filter is the discrete-time system output ${\bi X}$ and the filter output is an approximation to the discrete-time dynamic measurand ${\bi Y}$~\cite{EicElsEswHes10}. Due to the fact that this comprises an ill-posed inverse problem, some kind or regularization is necessary in order to obtain a meaningful estimate~\cite{Eicetal2016}. In general, two types of digital filters exist: finite impulse response (FIR) and infinite impulse response (IIR) filters. Most FIR filters have a linear phase response, thus do not cause non-linear distortions to the input signal. However, larger filter orders and thus longer time delays than for IIR filters may be necessary in order to achieve the desired filter behavior~\cite{OppSch89}. 

For an IIR filter an explicit measurement model is given by
\begin{equation}
Y(nT_s) = \sum_{k=0}^{N_b}\Theta_k X\left((n-k)T_s\right) - \sum_{k=1}^{N_a} \Phi_k Y\left( (n-k)T_s\right) + \Delta(nT_s)\, ,
\label{deconvolution filter}
\end{equation}
where $T_s$ denotes the length of the sampling interval. The filter coefficients ${\bi \Phi}=(\Phi_1,\ldots,\Phi_{N_a})^\top$ and ${\bi \Theta}=(\Theta_0,\ldots,\Theta_{N_b})^\top$ are determined from knowledge about the system model using, for instance, methods such as least-squares approximation~\cite{EicElsEswHes10}. The additional term $\Delta(nT_s)$ on the right-hand side of expression~(\ref{deconvolution filter}) denotes the correction for a time-varying error caused by the employed deconvolution filter~\cite{Eicetal2016}. For an FIR filter the coefficients ${\bi \Theta}$ in equation (\ref{deconvolution filter}) are equal to zero and thus the filter does not have a recursive part.

Assume that for a system with transfer function $H({\rm j}\omega)$ an $N$th order finite-impulse-response (FIR) filter with filter coefficients ${\bi \Theta}=(\Theta_0,\ldots,\Theta_M)$ is designed such that 
\[
H({\rm j}\omega)F_{\scriptsize \bi \Theta}(e^{{\rm j}\omega/T_s}) \approx \left\{ \begin{array}{ll} 1, & \textnormal{for } \omega\leq\bar\omega_1, \\ 0, & \textnormal{for } \omega \geq \bar\omega_2, \end{array}\right.
\]
where $F_{\scriptsize \bi \Theta}(e^{{\rm j}\omega/T_s})$ denotes the frequency response of the digital deconvolution filter and $\bar\omega_1\leq\bar\omega_2$ are chosen frequency values. Thus, the digital filter approximates the inverse of the system model in the frequency interval $[0,\bar\omega_1]$ and attenuates frequency components for $\omega\geq\bar\omega_2$. The attenuation of high-frequency components is necessary in order to avoid otherwise strong noise amplification. Therefore, the digital filter $F_{\scriptsize \bi \Theta}$ is decomposed into an (approximate) inverse filter $F_{\scriptsize \bi \Theta}^{\rm (inv)}$ and a low-pass filter $F_{\scriptsize \bi \Theta}^{\rm (low)}$, both of FIR type. The (approximate) inverse filter $F_{\scriptsize \bi \Theta}^{\rm (inv)}$ satisfies $H({\rm j}\omega)F_{\scriptsize \bi \Theta}^{\rm (inv)}\approx 1$ for $\omega\leq \bar\omega_1$, and typically requires to introduce a time-delay $\tau_0=n_0T_s$ as a kind of regularization~\cite{ParBit87}. The low-pass filter part $F_{\scriptsize \bi \Theta}^{\rm (low)}$ satisfies $F_{\scriptsize \bi \Theta}^{\rm (low)}(e^{{\rm j}\omega/T_s})\approx 1$ for $\omega\leq \omega_1$ and realizes the desired attenuation at frequencies $\omega\geq\bar\omega_2$. Both filter parts are applied in cascade to obtain an estimate of the dynamic measurand. 

The choice of the frequency bounds $\bar\omega_1$ and $\bar\omega_2$ depends on the available knowledge or assumptions about the frequency content of the measurand. When the bandwidth (largest significant frequency) of the measurand is smaller than or equal to $\bar\omega_1$, then the application of the deconvolution filter is a correct model for the measurand. Otherwise time-varying errors are introduced.

The decomposition of the deconvolution filter into an (approximate) inverse of the system model and a low-pass filter is a typical approach to deconvolution~\cite{Eic12}. The (approximate) inverse part can be determined by regression as in the above example, or by other means~\cite{EicElsEswHes10}. For the application of an FIR filter as a measurement model, the evaluation of uncertainty associated with an estimate of the measurand can be carried out analytically~\cite{ElsLin08}. When the measurement model is an infinite-impulse response (IIR) filter, either Monte Carlo uncertainty propagation~\cite{EicLinHarEls12} is carried out or the measurement model is transformed into a state-space representation~\cite{OppSch89}
\[
\begin{array}{lcl}
	{\bi z}[n] &=& {\bi C}{\bi z}[n-1] + {\bi D}{\bi x}[n-1] \\
	{\bi y}[n] &=& {\bi E}{\bi z}[n]   + {\bi F}{\bi x}[n]   ,
\end{array}
\]
allowing for a direct application of the law of propagation of uncertainty of GUM \cite{LinEl09}.

The additional low-pass filter, required to attenuate high-frequency noise, introduces an error to the measurement model that can usually not be avoided and contributes to the uncertainty associated with the measurand~\cite{Eic12}. The reason for this error is that the low-pass filter part of the deconvolution filter also attenuates high-frequency components of the estimate of the measurand. The derivation of the model is thus typically a trade-off between the attenuation of variance due to measurement noise and the reduction of estimation errors. 

\section{PyDynamic: background and use}

To implement the methods described above we have developed an open source software package called PyDynamic that is available from the Github repository (\url{https://github.com/eichstaedtPTB/PyDynamic}). It provides a range of data analysis tools that are suitable for dynamic measurements.  The repository contains a detailed description of the software, with instructions on how to install and use it.   

PyDynamic, which is written in the Python programming language version 3.5, offers propagation of uncertainties for:
\begin{itemize}
    \item application of the Discrete Fourier Transform and its inverse;
    \item filtering with an FIR or IIR filter with uncertain coefficients;
    \item design of an FIR filter as the inverse of a frequency response with uncertain coefficients;
    \item design of an IIR filter as the inverse of a frequency response with uncertain coefficients;
    \item deconvolution in the frequency domain by division;
    \item multiplication in the frequency domain;
    \item transformation from amplitude and phase to a representation by real and imaginary parts.
\end{itemize}

A useful account of the background to the PyDynamic software and its main functions can be found in~\cite{EiElSmTe17}. The main idea of the PyDynamic package is to provide a comprehensive toolset for the analysis of dynamic measurements in which the propagation of uncertainties is taken care of as easily as possible for the user. 

The available tools include routines designed for second order systems such as accelerometers: calculation of the system's frequency response, calculation of continuous filter coefficients from physical parameters, propagation of uncertainty from physical parameters to the complex system's transfer function (either defined in terms of real and imaginary components or amplitude and phase components) using  Monte Carlo methods as described in GUM Supplement 2~\cite{JCGM102}.  There is also a range of filter design tools including calculation of the group delay of a digital filter, design of an FIR low pass filter using the window technique with a Kaiser window, determining whether an IIR filter is stable, and smoothing (and optionally differentiating) data with a Savitzky-Golay filter. The software also provides tools for generating test signals that can be used to investigate and validate signal processing routines developed either within PyDynamic or externally. The relevant routines perform the following tasks: generate a signal that resembles a shock excitation as a Gaussian followed by a smaller Gaussian of opposite sign, generate a Gaussian pulse at $t(0)$ with height $m(0)$ and standard deviation $\sigma$, generate a rectangular signal of given height and width $t(1)-t(0)$, and generate a series of rectangular functions to represent a square pulse signal. 

\section{Example applications}

This section describes four examples analysed using PyDynamic routines in order to provide a good starting point for end-users. All examples have been published in detail elsewhere and are here presented in a concise way to illustrate the use of the PyDynamic software package.

\subsection{Shock calibration}
\begin{figure}[htp]
\centerline{\includegraphics[height=80mm,width=120mm]{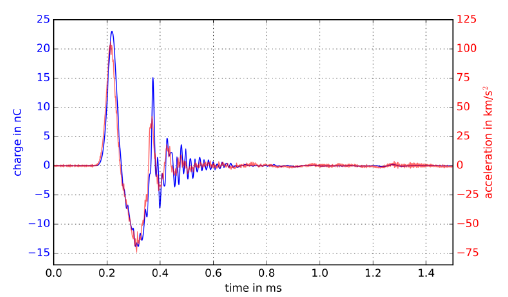}}
\caption{\small Example of a set of shock measurement time series, input acceleration (red) and output charge (blue).}
\label{fig: Bruns shock}
\end{figure}

A typical workflow for a deconvolution problem using PyDynamic is described below. The problem of interest is the calibration of an accelerometer using a high-intensity shock, and is described in detail in~\cite{Bruns2017}. After data acquisition using LabVIEW~\cite{LabVIEW} and data pre-processing, the problem of interest is as shown in Figure~\ref{fig: Bruns shock}, where the red line shows the input acceleration time series, and the blue line shows the output time series from a charge amplifier.

The accelerometer is modelled as an LTI second-order system and for the purpose of the calibration the system model is written in the following form: 
\begin{equation}
S(\omega)=S_0\frac{\omega^{2}_{0}}{\omega^{2}_{0}+2{\rm j}\omega\delta\omega_{0}-\omega^{2}}.
\label{Bruns-equation1}
\end{equation}
To permit frequency domain analysis, the two time series from Figure~\ref{fig: Bruns shock} were Fourier transformed using the \texttt{GUM\_DFT} routine from PyDynamic. This takes a time series and measurement uncertainties associated with each data point of the series and outputs the complex Fourier coefficients and the complete covariance matrix for the frequency domain results. See~\cite{EiWi16} for more information on the GUM2DFT routines included in PyDynamic.

The next stage of the process is to take into account and correct for the effect of the measuring chain on the output time series, where it is necessary to convert the voltage output from the measuring system to the charge output of the accelerometer. The equation that describes the charge sensitivity as a function of frequency is:
\[
S_{qs}(\omega)=\frac{F_{v}(\omega)}{S_{uq}(\omega)}.\frac{1}{F_{a}(\omega)}=\frac{F_q(\omega)}{F_{a}(\omega)},
\]
where $F_{a}$ and $F_{v}$ are the Fourier coefficients of, respectively, the acceleration and the measuring chain voltage, and $q$ indicates charge. Note that division of a frequency response by another frequency response can be regarded as a deconvolution operation and can be performed using the DFT deconvolution routine \texttt{DFT\_deconv} from PyDynamic.

\begin{figure}[hbt]
\centering
  \includegraphics[width=0.8\columnwidth]{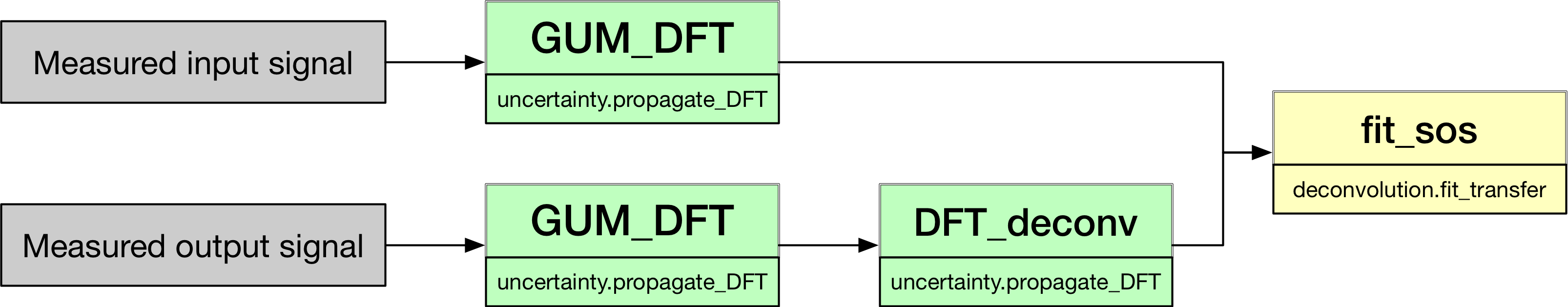}
  \caption{Flow chart of using PyDynamic for the parametric calibration of an accelerometer using time domain measurements. In each step, measurement uncertainties are evaluated by the PyDynamic routines.}
  \label{fig:flow chart shock accel}
\end{figure}

To carry out the parameter identification required for equation~(\ref{Bruns-equation1}), re-arrangement and normalization produces:
\begin{equation}
\frac{S_{0}}{S(\omega)}= 1 + 2{\rm j}\frac{\delta}{\omega_{0}}\omega - \frac{1}{\omega^{2}_{0}}\omega^{2}.
\end{equation}
The PyDynamic routine \texttt{fit\_sos} can be adapted to solve this equation. Figure~\ref{fig:flow chart shock accel} outlines the complete use of PyDynamic routines to solve the task of parametric calibration starting from the time domain measurements.

To test the methodology and the use of PyDynamic for parameter identification the authors of~\cite{Bruns2017} used a Monte Carlo method. A digital IIR filter was designed, using PyDynamic routines, that could be regarded as a software implementation of a calibrated accelerometer, so that given an acceleration time series as an input it would return the same charge output time series as the accelerometer. Values of the set of parameters $S_{0}, \delta$ and $\omega_{0}$ were drawn randomly 2\,000 times from a joint multivariate normal distribution. For each set of values an IIR filter was designed using the following procedure:
\begin{enumerate}
\item Obtain the coefficients of an analogue filter to represent the second order system using the PyDynamic second order system tools for each of the $N = 2\,000$ sets;
\item Using {\em SciPy} routines~\cite{SciPy} to obtain the coefficients of a digital filter for a given sampling rate;
\item Apply the digital filter to the acceleration time series;
\item Update a time series of the cumulative mean and variance for each of the $N$ time series of charge;
\item Compare the original measurement with the calculated charge time series and its bounds, as determined from the Monte Carlo calculation.
\end{enumerate}

\subsection{Piezoelectric fiber optic sensor}
\begin{figure}[htp]
\centerline{\includegraphics[height=80mm,width=120mm]{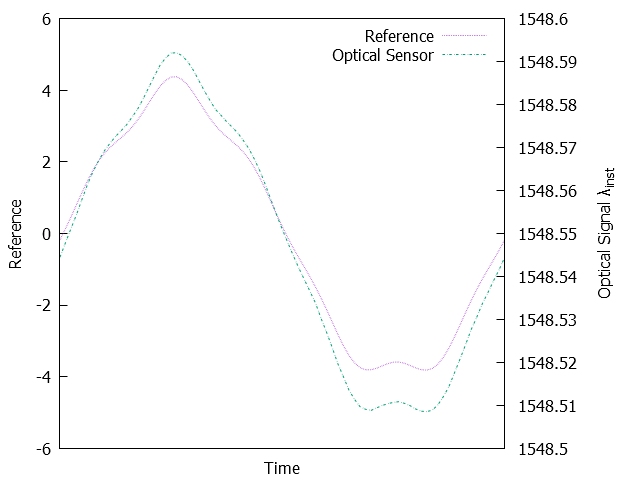}}
\caption{\small Example of a sensor measurement with corresponding reference equivalent.}
\label{fig:piezo1}
\end{figure}

The problem of interest is to compensate the hysteresis effects of a piezoelectric fiber optic sensor~\cite{Fusi05,Fusi07}. The sensors are measuring approximately sinusoidal properties, though some distortion may be present. An example of a sensor measurement and corresponding reference measurement in the time domain is shown in Figure~\ref{fig:piezo1}. 

The processing that is applied takes the reference and sensor measurements, transforms them into the frequency domain, and then divides the first by the second in the frequency domain to give the system transfer function. Any future measurement by the sensor is transformed into the frequency domain, multiplied by the transfer function and the result is transformed back into the time domain resulting in a compensated time series. The \texttt{DFT\_transferfunction} routine from PyDynamic was used to undertake the calculation of the transfer function together with an associated covariance matrix as an indicator of its uncertainty.

\begin{figure}[hbt]
  \centering
  \includegraphics[width=0.8\columnwidth]{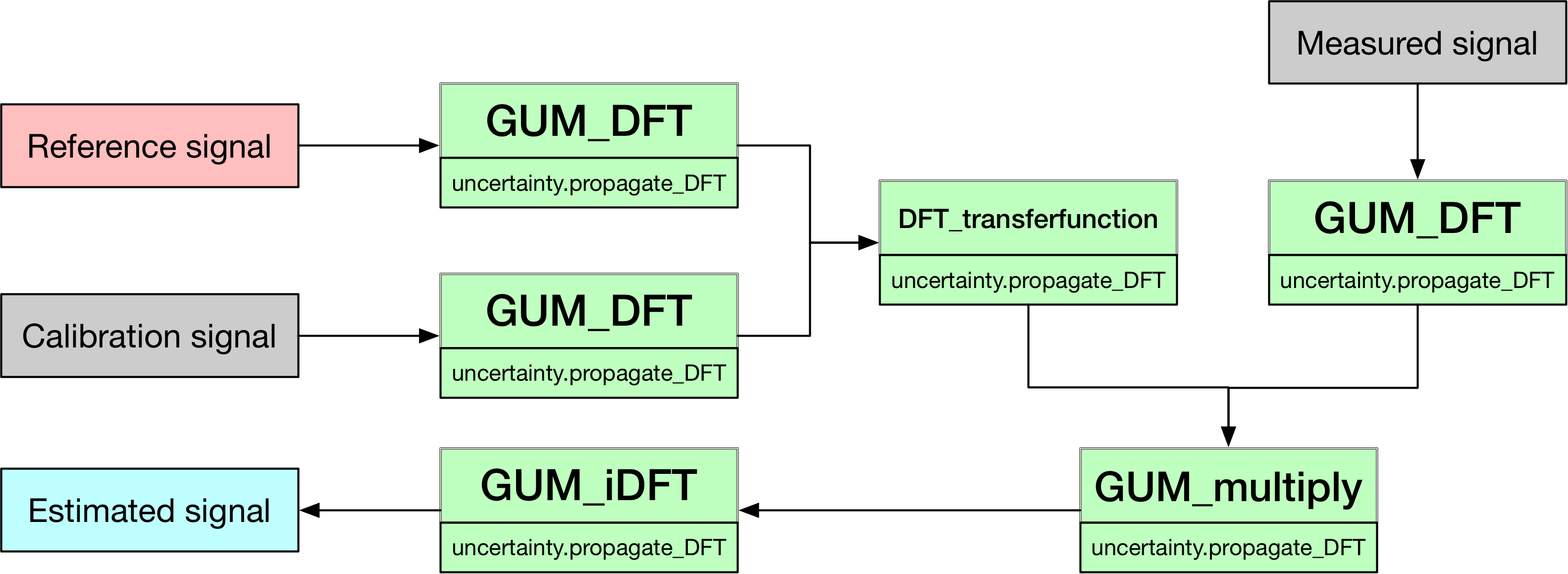}
  \caption{Flow chart of applying the PyDynamic routines for the fiber optic sensor example. Measurement uncertainties are evaluated in each step by the respective PyDynamic routines.}
  \label{fig: flow chart for fiber optic sensor}
\end{figure}

\begin{figure}[htp]
\centerline{\includegraphics[height=80mm,width=120mm]{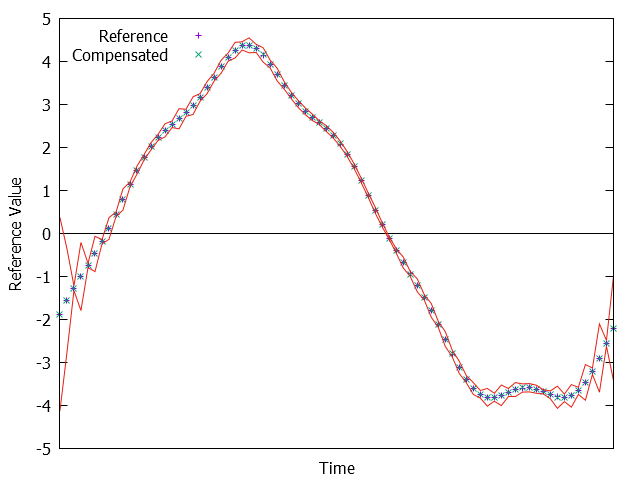}}
\caption{\small Comparison of reference signal against compensated signal with associated uncertainties.}
\label{fig:piezo2}
\end{figure}

New measurements are transformed into the frequency domain using the \texttt{GUM\_DFT} routine, and the transfer function is applied using the \texttt{DFT\_multiply} routine to perform the compensation. The final step of converting back into the time domain is undertaken using the \texttt{GUM\_iDFT} routine giving an estimate and uncertainty for each time instant. Figure~\ref{fig: flow chart for fiber optic sensor} shows the complete workflow using the PyDynamic routines for this example. The final result is shown in Figure ~\ref{fig:piezo2}, where the reference time signal is compared against the compensated time series and its associated uncertainties. 

As shown in the example, the signals do not contain many frequencies, and the creation of transfer functions with associated uncertainties can be a problem when only noise is measured. Such issues arising from the application of PyDynamic would have occurred with any frequency domain analysis. However, PyDynamic's functionality to propagate uncertainties significantly helped to show when only noise was being measured. Note that the transfer function was not constant with the peak measurement of the sensor, which required transfer functions to be calculated for a range of peak values and a fitting tool used to correlate peak values to components of the transfer function. 

\subsection{Hydrophone deconvolution}
For the study of medical ultrasound devices, measurements with calibrated hydrophones are carried out for specific test signals. In the corresponding data analysis, the unwanted effects from the hydrophone have to be compensated based on the available calibration information. The measured ultrasound signals typically show a strong time dependence with bandwidth close to that of the hydrophone used for the measurement. Thus, a deconvolution is advocated in order to reduce the time-varying deviations caused by the measuring instrument. This approach is of particular importance when the analysis of the measured ultrasound signal is used to validate that the device is working within specified limits in order to avoid operations harmful for the patient. Therefore, negative and positive peak values are to be determined. Improved measurement of these values allows to reduce safety margins otherwise necessary to meet regulatory requirements.

\begin{figure}[hbt]
\centering
  \includegraphics[width=0.8\columnwidth]{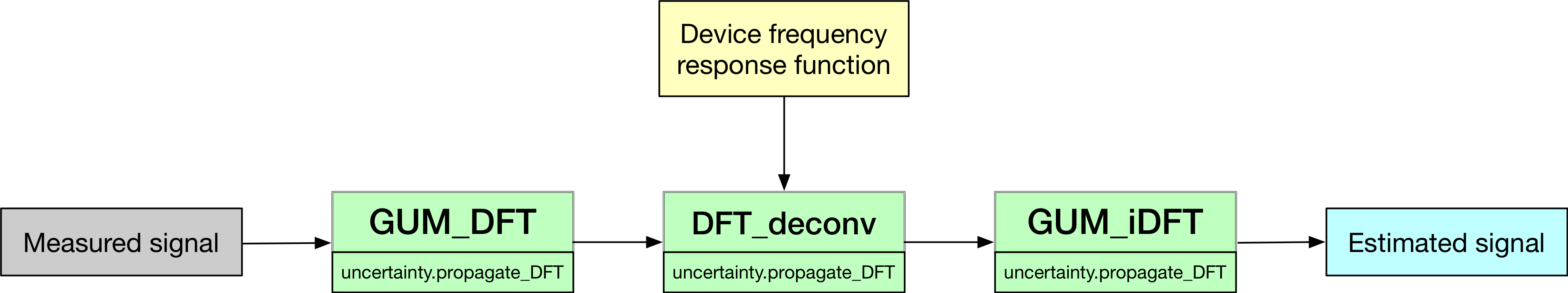}
  \caption{Flow chart of applying the PyDynamic routines for the deconvolution of hydrophone measurements. Uncertainties associated with the measurement and with the calibration data of the frequency response function are taken into account by the PyDynamic routines.}
  \label{fig: flow chart hydrophone}
\end{figure}

In \cite{EiWi16} the authors have studied a generic approach for the deconvolution of hydrophone measurements provided the complex-valued frequency response function $H(f)$ of the measurement device is available at equidistant frequency points over the full range of frequencies (from 0~Hz to the Nyquist frequency). Then, the estimate $\widehat{p}(t)$ of the ultrasound pressure signal is estimated from the hydrophone output signal $x(t)$ as follows:
\begin{enumerate}
\item Transform the hydrophone output signal to the frequency domain by application of the Discrete Fourier Transform (DFT): \[ X(f) = \mathcal{F}(x(t)); \]
\item Multiply the reciprocal of the frequency response function and the frequency response of a chosen low-pass filter: \[ \widehat{P}(f) = X(f) H^{-1}(f) H_{\rm low}(f); \]
\item Transform the result back to the time domain using the inverse Fourier transform: \[ \widehat{p}(t) = \mathcal{F}^{-1}(\widehat{P}(f)). \]
\end{enumerate}
The low-pass filter is necessary in order to suppress high-frequency noise which is amplified due to the application of the reciprocal frequency response function $H^{-1}(f)$. The complete workflow in PyDynamic for this example is shown in Figure~\ref{fig: flow chart hydrophone}.

The application of the low-pass filter causes a systematic error -- the so-called regularization error. Whenever this error is significant in size compared to the other uncertainty sources, it has to be accounted for in the uncertainty budget. Therefore, in \cite{EiWi17} a simplified method for the evaluation of the regularization error was proposed. It is planned to incorporate this method into PyDynamic in the near future.

\subsection{Blood pressure analysis}
In \cite{Ogor18} the authors study invasive blood pressure (IBP) measurements, which is an important medical measurement method in intensive health care in hospitals. It is well-known that the  quality of the measured signal depends on the tubing system from the measurement tip to the data acquisition system. Moreover, it is necessary to verify the reliability of typical measurement setups under dynamic conditions.

\begin{figure}[hbt]
\centering
  \includegraphics[width=0.9\columnwidth]{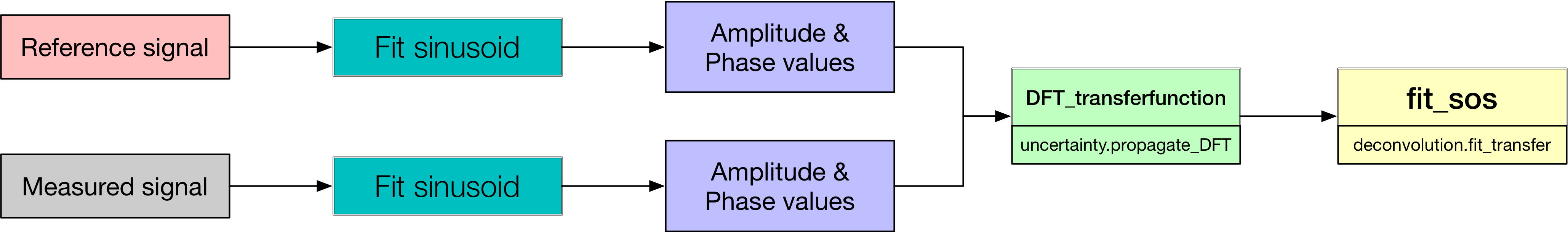}
  \caption{Flow chart of applying the PyDynamic routines for the sinusoidal calibration of an invasive blood pressure device using a pressure generator and a reference pressure transducer.}
  \label{fig:flow chart IBP calibration}
\end{figure}

In a first step, a calibration setup for IBP measurements was developed using a reference pressure transducer and a pressure generator. The generator provided single-frequency sinusoidal pressure signals. By changing the frequency of the sinusoid, the system's frequency response over the relevant frequency band from 0~Hz to 25~Hz was evaluated with frequency-wise uncertainties in amplitude and phase. A second-order system can be fitted to the obtained frequency response values using the \texttt{fit\_sos} routine from PyDynamic resulting in the calibration workflow shown in Figure~\ref{fig:flow chart IBP calibration}. For a future revision of PyDynamic it is planned to also implement methods for fitting sinusoidal data.

The fitted transfer function model allows to interpolate between the frequencies from the calibration and to extrapolate beyond the upper frequency limit from the calibration. The authors in \cite{Ogor18} used this information for the design of a suitable deconvolution filter. That is, the authors designed a digital filter which compensates for the dynamic characteristics of the measurement device, see \cite{OppSch89, Eic12}. In PyDynamic the corresponding routine is \texttt{LSFIR} for a finite impulse response (FIR) filter and \texttt{LSIIR} for an infinite impulse response (IIR) filter, see~\cite{OppSch89}.

\begin{figure}[hbt]
\centering
  \includegraphics[width=0.9\columnwidth]{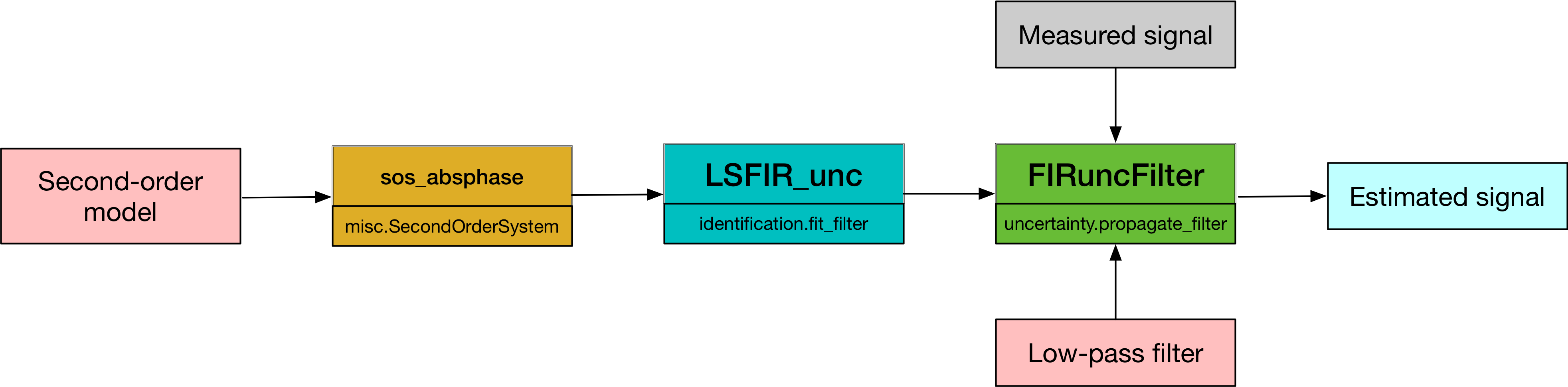}
  \caption{Flow chart of applying PyDynamic routines for the deconvolution of invasive blood pressure measurements. In each step measurement uncertainties are propagated by the PyDynamic routines.}
  \label{fig:flow chart IBP measurement}
\end{figure}

With the PyDynamic routines \texttt{SMC} for a memory-efficient implementation of the Monte Carlo method for dynamic measurements \cite{EicLinHarEls12} and \texttt{FIRuncFilter} for a closed-formula approach for FIR filters, the authors in \cite{Ogor18} were able to test easily the application of the original GUM and its Supplement 1 methods. The analysis showed a very good agreement of both approaches as expected for FIR filters. Figure~\ref{fig:flow chart IBP measurement} shows the workflow for the application of an FIR deconvolution filter and the uncertainty propagation formula~\cite{ElsLin08}.

\section{Summary and outlook}
The analysis and characterization of dynamic measurements is a topic of growing importance and a large amount of literature exists for the mathematical modelling and for the evaluation of uncertainties. The challenge for end users, though, is that this field joins diverse elements from measurement science, statistics, mathematics and signal processing. This contribution outlined the basic principles of dynamic measurement analysis methods, which are necessary for end users to understand the concept and application of available methods. With the open-source software library PyDynamic, end users are further supported in the application of the mathematical and statistical methods for uncertainty evaluation in dynamic measurements. Actual examples from measurement science showed the utilization of PyDynamic and can be used as starting point for other examples. Based on user feedback and research at the collaborating NMIs, the functionality of PyDynamic is expanded continuously.

\section*{Acknowledgement}
This work is part of the Joint Support for Impact project 14SIP08 of the European Metrology Programme for Innovation and Research (EMPIR). The EMPIR is jointly funded by the EMPIR participating countries within EURAMET and the European Union.

\end{document}